\documentclass[%
 aip,
cp,  % Conference Proceedings
 amsmath,amssymb,%nobibnotes,
 reprint,%
]{revtex4-2}

\usepackage{graphicx}% Include figure files
\usepackage{dcolumn}% Align table columns on decimal point
\usepackage{bm}% bold math
\usepackage[utf8]{inputenc}
\usepackage[T1]{fontenc}
\usepackage{mathptmx} 
\usepackage{mhchem}
\usepackage{nccmath}

\def\MJ{{\textsc{Majorana }}}
\def\MJD{{\textsc{Majorana Demonstrator }}}
\def\MJDnospace{{\textsc{Majorana Demonstrator}}}
\def\DEM{{\textsc{Demonstrator }}}
\def\DEMnospace{{\textsc{Demonstrator}}}
\def\EnrGe{\ce{^{76}Ge} }
\def\EnrGenospace{\ce{^{76}Ge}}
\def\ndbd{$0\nu\beta\beta$ }
\def\ndbdnospace{$0\nu\beta\beta$}

\def\Thorium{\ce{^{232}Th }}

\def\Uraniumns{\ce{^{238}U}}
\def\Thoriumns{\ce{^{232}Th}}
\def\Potassiumns{\ce{^{40}K}}
\def\CosmoGe{\ce{^{68}Ge}}
\def\CosmoCo{\ce{^{60}Co}}
\def\CosmoCotwo{\ce{^{57}Co}}

\def\Uranonns{\ce{^{222}Rn}}

\def\Thallium{\ce{^{208}Tl} }

\def\Bismuthns{\ce{^{214}Bi}}

\def\Lead{\ce{^{210}Pb}}

\begin{document}

%\title{Modeling Backgrounds for the \MJDnospace{}}% Force line breaks with \\
\title{Modeling Backgrounds for the \textbf{\textsc{Majorana Demonstrator}}}% Force line breaks with \\

\author{C.R. Haufe} % Write as First name Surname
 \email[Corresponding author: ]{crhaufe@live.unc.edu}
\affiliation{Department of Physics and Astronomy, University of North Carolina, Chapel Hill, NC, USA}
\affiliation{Triangle Universities Nuclear Laboratory, Durham, NC, USA}

\author{I.J. Arnquist}
\affiliation{Pacific Northwest National Laboratory, Richland, WA, USA}

\author{F.T. Avignone III}
\affiliation{Department of Physics and Astronomy, University of South Carolina, Columbia, SC, USA}
\affiliation{Oak Ridge National Laboratory, Oak Ridge, TN, USA}

\author{A.S. Barabash}
\affiliation{National Research Center ``Kurchatov Institute” Institute for Theoretical and Experimental Physics, Moscow, Russia}

\author{\\C.J. Barton}
\affiliation{Department of Physics, University of South Dakota, Vermillion, SD, USA}

\author{K.H. Bhimani}
\affiliation{Department of Physics and Astronomy, University of North Carolina, Chapel Hill, NC, USA}
\affiliation{Triangle Universities Nuclear Laboratory, Durham, NC, USA}

\author{E. Blalock}
\affiliation{Department of Physics, North Carolina State University, Raleigh, NC, USA}
\affiliation{Triangle Universities Nuclear Laboratory, Durham, NC, USA}

\author{B. Bos}
\affiliation{Department of Physics and Astronomy, University of North Carolina, Chapel Hill, NC, USA}
\affiliation{Triangle Universities Nuclear Laboratory, Durham, NC, USA}

\author{M. Busch}
\affiliation{Department of Physics, Duke University, Durham, NC, USA}
\affiliation{Triangle Universities Nuclear Laboratory, Durham, NC, USA}

\author{\\M. Buuck}
\affiliation{Center for Experimental Nuclear Physics and Astrophysics, and Department of Physics,\\ University of Washington, Seattle, WA, USA}

\author{T.S. Caldwell}
\affiliation{Department of Physics and Astronomy, University of North Carolina, Chapel Hill, NC, USA}
\affiliation{Triangle Universities Nuclear Laboratory, Durham, NC, USA}

\author{Y-D. Chan}
\affiliation{Nuclear Science Division, Lawrence Berkeley National Laboratory, Berkeley, CA, USA}

\author{C.D. Christofferson}
\affiliation{South Dakota Mines, Rapid City, SD, USA}

\author{P.-H. Chu}
\affiliation{Los Alamos National Laboratory, Los Alamos, NM, USA}

\author{M.L. Clark}
\affiliation{Department of Physics and Astronomy, University of North Carolina, Chapel Hill, NC, USA}
\affiliation{Triangle Universities Nuclear Laboratory, Durham, NC, USA}

\author{C. Cuesta}
\affiliation{Centro de Investigaciones Energ\'eticas, Medioambientales y Tecnol\'ogicas, CIEMAT 28040, Madrid, Spain}

\author{J.A. Detwiler}
\affiliation{Center for Experimental Nuclear Physics and Astrophysics, and Department of Physics,\\ University of Washington, Seattle, WA, USA}

\author{Yu. Efremenko}
\affiliation{Department of Physics and Astronomy, University of Tennessee, Knoxville, TN, USA}
\affiliation{Oak Ridge National Laboratory, Oak Ridge, TN, USA}

\author{H. Ejiri}
\affiliation{Research Center for Nuclear Physics, Osaka University, Ibaraki, Osaka, Japan}

\author{\\S.R. Elliott}
\affiliation{Los Alamos National Laboratory, Los Alamos, NM, USA}

\author{G.K. Giovanetti}
\affiliation{Physics Department, Williams College, Williamstown, MA, USA}

\author{M.P. Green}
\affiliation{Department of Physics, North Carolina State University, Raleigh, NC, USA}
\affiliation{Triangle Universities Nuclear Laboratory, Durham, NC, USA}
\affiliation{Oak Ridge National Laboratory, Oak Ridge, TN, USA}

\author{J. Gruszko}
\affiliation{Department of Physics and Astronomy, University of North Carolina, Chapel Hill, NC, USA}
\affiliation{Triangle Universities Nuclear Laboratory, Durham, NC, USA}

\author{I.S. Guinn}
\affiliation{Department of Physics and Astronomy, University of North Carolina, Chapel Hill, NC, USA}
\affiliation{Triangle Universities Nuclear Laboratory, Durham, NC, USA}

\author{V.E. Guiseppe}
\affiliation{Oak Ridge National Laboratory, Oak Ridge, TN, USA}

\author{R. Henning}
\affiliation{Department of Physics and Astronomy, University of North Carolina, Chapel Hill, NC, USA}
\affiliation{Triangle Universities Nuclear Laboratory, Durham, NC, USA}

\author{D. Hervas Aguilar}
\affiliation{Department of Physics and Astronomy, University of North Carolina, Chapel Hill, NC, USA}
\affiliation{Triangle Universities Nuclear Laboratory, Durham, NC, USA}

\author{E.W. Hoppe}
\affiliation{Pacific Northwest National Laboratory, Richland, WA, USA}

\author{\\A. Hostiuc}
\affiliation{Center for Experimental Nuclear Physics and Astrophysics, and Department of Physics,\\ University of Washington, Seattle, WA, USA}

\author{M.F. Kidd}
\affiliation{Tennessee Tech University, Cookeville, TN, USA}

\author{I. Kim}
\affiliation{Los Alamos National Laboratory, Los Alamos, NM, USA}

\author{R.T. Kouzes}
\affiliation{Pacific Northwest National Laboratory, Richland, WA, USA}

\author{T.E. Lannen V}
\affiliation{Department of Physics and Astronomy, University of South Carolina, Columbia, SC, USA}

\author{A. Li}
\affiliation{Department of Physics and Astronomy, University of North Carolina, Chapel Hill, NC, USA}
\affiliation{Triangle Universities Nuclear Laboratory, Durham, NC, USA}

\author{A.M. Lopez}
\affiliation{Department of Physics and Astronomy, University of Tennessee, Knoxville, TN, USA}

\author{J.M. L\'opez-Casta\~no}
\affiliation{Oak Ridge National Laboratory, Oak Ridge, TN, USA}

\author{E.L. Martin}
\affiliation{Department of Physics and Astronomy, University of North Carolina, Chapel Hill, NC, USA}
\affiliation{Triangle Universities Nuclear Laboratory, Durham, NC, USA}

\author{R.D. Martin}
\affiliation{Department of Physics, Engineering Physics and Astronomy, Queen’s University, Kingston, ON, Canada}

\author{\\R. Massarczyk}
\affiliation{Los Alamos National Laboratory, Los Alamos, NM, USA}

\author{S.J. Meijer}
\affiliation{Los Alamos National Laboratory, Los Alamos, NM, USA}

\author{T.K. Oli}
\affiliation{Department of Physics, University of South Dakota, Vermillion, SD, USA}

\author{G. Othman}
\affiliation{Department of Physics and Astronomy, University of North Carolina, Chapel Hill, NC, USA}
\affiliation{Triangle Universities Nuclear Laboratory, Durham, NC, USA}

\author{L.S. Paudel}
\affiliation{Department of Physics, University of South Dakota, Vermillion, SD, USA}

\author{\\W. Pettus}
\affiliation{Department of Physics, Indiana University, Bloomington, IN, USA}
\affiliation{IU Center for Exploration of Energy and Matter, Bloomington, IN, USA}

\author{A.W.P. Poon}
\affiliation{Nuclear Science Division, Lawrence Berkeley National Laboratory, Berkeley, CA, USA}

\author{D.C. Radford}
\affiliation{Oak Ridge National Laboratory, Oak Ridge, TN, USA}

\author{A.L. Reine}
\affiliation{Department of Physics and Astronomy, University of North Carolina, Chapel Hill, NC, USA}
\affiliation{Triangle Universities Nuclear Laboratory, Durham, NC, USA}

\author{K. Rielage}
\affiliation{Los Alamos National Laboratory, Los Alamos, NM, USA}

\author{\\N.W. Ruof}
\affiliation{Center for Experimental Nuclear Physics and Astrophysics, and Department of Physics,\\ University of Washington, Seattle, WA, USA}

\author{D.C. Schaper}
\affiliation{Los Alamos National Laboratory, Los Alamos, NM, USA}

\author{D. Tedeschi}
\affiliation{Department of Physics and Astronomy, University of South Carolina, Columbia, SC, USA}

\author{R.L. Varner}
\affiliation{Oak Ridge National Laboratory, Oak Ridge, TN, USA}

\author{S. Vasilyev}
\affiliation{Joint Institute for Nuclear Research, Dubna, Russia}

\author{\\J.F. Wilkerson}
\affiliation{Department of Physics and Astronomy, University of North Carolina, Chapel Hill, NC, USA}
\affiliation{Triangle Universities Nuclear Laboratory, Durham, NC, USA}
\affiliation{Oak Ridge National Laboratory, Oak Ridge, TN, USA}

\author{C. Wiseman}
\affiliation{Center for Experimental Nuclear Physics and Astrophysics, and Department of Physics,\\ University of Washington, Seattle, WA, USA}

\author{W. Xu}
\affiliation{Department of Physics, University of South Dakota, Vermillion, SD, USA}

\author{C.-H. Yu}
\affiliation{Oak Ridge National Laboratory, Oak Ridge, TN, USA}

\author{B.X. Zhu}
\affiliation{Los Alamos National Laboratory, Los Alamos, NM, USA}

\date{\today} % It is always \today, today, but any date may be explicitly specified
              % Not printed for conference proceedings

\begin{abstract}
The \MJD{}is a neutrinoless double-beta decay (\ndbdnospace{}) experiment containing $\sim$30 kg of p-type point-contact germanium detectors enriched to 88\% in \EnrGe{} and $\sim$14 kg of natural germanium detectors. The detectors are housed in two electroformed copper cryostats and surrounded by a graded passive shield with an active muon veto. An extensive radioassay campaign was performed prior to installation to insure the use of ultra-clean materials. The \DEM{} achieved one of the lowest background rates in the region of the \ndbd Q-value, 15.7 $\pm$ 1.4 cts/(FWHM t y) from the low-background configuration spanning most of the 64.5 kg-yr active exposure. Nevertheless this background rate is a factor of five higher than the projected background rate. This discrepancy arises from an excess of events from the \Thorium{} decay chain. Background-model fits aim to %understand this
explain the deviation from assay-based projections, potentially determine the source(s) of observed backgrounds, and allow a %precision 
precise measurement of the two-neutrino double-beta decay half-life. The fits agree with earlier simulation studies, which indicate the origin of the \Thorium{} excess is not from a near-detector component and have informed design decisions for the next-generation LEGEND experiment. Recent findings have narrowed the suspected locations for the excess activity, motivating a final simulation and assay campaign to complete the background model.
\end{abstract}

\maketitle

\section{\label{sec:intro}INTRODUCTION}

Neutrinoless double beta decay is a lepton-number-violating process, the discovery of which would imply the Majorana nature of neutrinos.  The \MJD is an effort to demonstrate the application of techniques and technologies that would allow for a next-generation experiment to probe a \ndbd half-life up to $10^{28}$ years.  Sensitivity to this half-life would cover the neutrino mass scale in the inverted neutrino mass ordering region.  In order to achieve this sensitivity, background radiation is heavily suppressed through mitigation techniques.  The \DEM achieved the second lowest background of any \ndbd experiment at $Q_{\beta\beta}$ (the Q value of both \ndbd and $2\nu\beta\beta$) through the use of ultra-clean materials, a graded passive shield, and an active muon veto.  Many of these technologies will be deployed in the next-generation LEGEND experiment.

\subsection{\label{sec:intromjd}\MJD Overview}
The \DEM operated at the 4850' level of the Sanford Underground Research Facility in Lead, SD, USA.  It consisted of two arrays of detectors (referred here as Module 1 and Module 2), each housed in electroformed copper cryostats that were machined on-site underground.  These detectors include $\sim$30 kg of p-type point contact detectors enriched to 88\% \EnrGenospace{}, $\sim$14 kg of BEGe detectors with the natural abundance of \EnrGenospace{}, and in later datasets 6.7 kg of enriched-\EnrGe{} inverted coaxial point-contact detectors (see \cite{znbb22} for additional details on the detector geometries and configuration during the operation of the \DEMnospace{}).  The benefit of using p-type point-contact detectors is their excellent energy resolution, measured in the \DEM to be 2.5 keV FWHM at the $Q_{\beta\beta}$ of 2039 keV.

\begin{figure}
    \centering
    \includegraphics[scale=0.4]{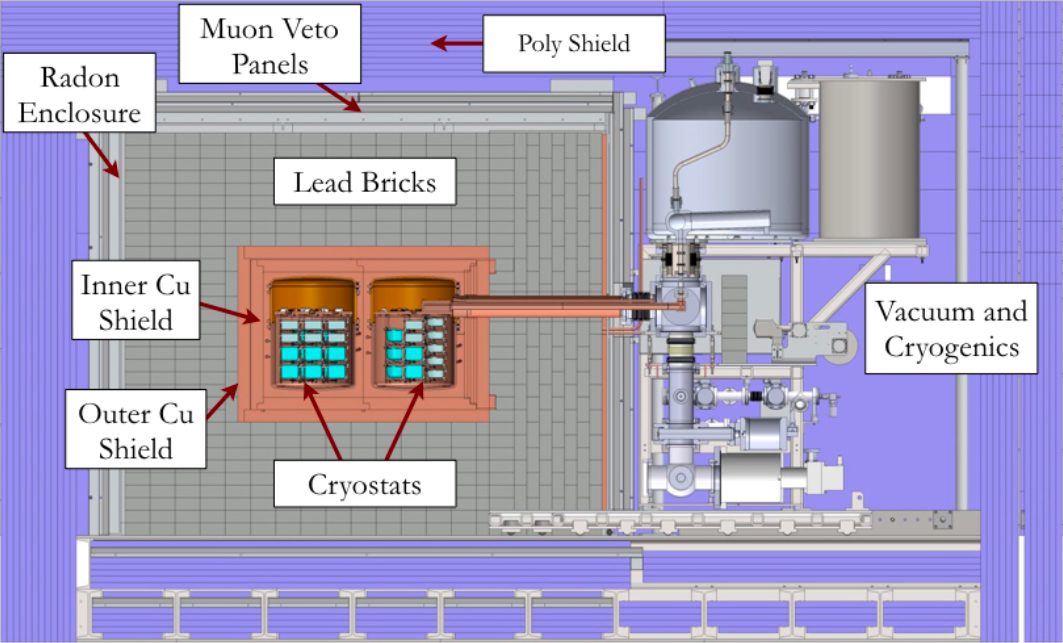}
    \caption{Shield layout of the \MJDnospace{}. The crossarm, vacuum system, and cryogenic systems are  shown  for only one module.}
    \label{fig:apparatus}
\end{figure}

The two modules are contained within several layers of shielding.  From the detector array out towards the cavern walls, this includes a layer of electroformed copper (Inner Cu Shield), a layer of commercial copper (Outer Cu Shield), a layer of lead bricks, a radon-exclusion enclosure purged with nitrogen gas, a layer of scintillation panels that form an active muon veto, and finally several layers of borated polyethylene.  Two hollow electroformed copper tubes, referred to as crossarms, are installed within the shielding to connect each module interior to %their
its respective vacuum and cryogenic system (see Figure \ref{fig:apparatus}).

In order to validate the radioactivity of the ultra-clean components used in the \DEMnospace{}, an extensive radioassay campaign was carried out prior to installation \cite{radioassay}.  The results of this radioassay campaign formed the basis of the assay-based background model.  This model combines the measured assay value of each component with the simulated efficiencies for each component to produce a predicted comprehensive background energy spectrum for the \DEMnospace{}.

\begin{figure}
    \centering
    \includegraphics[width=\textwidth]{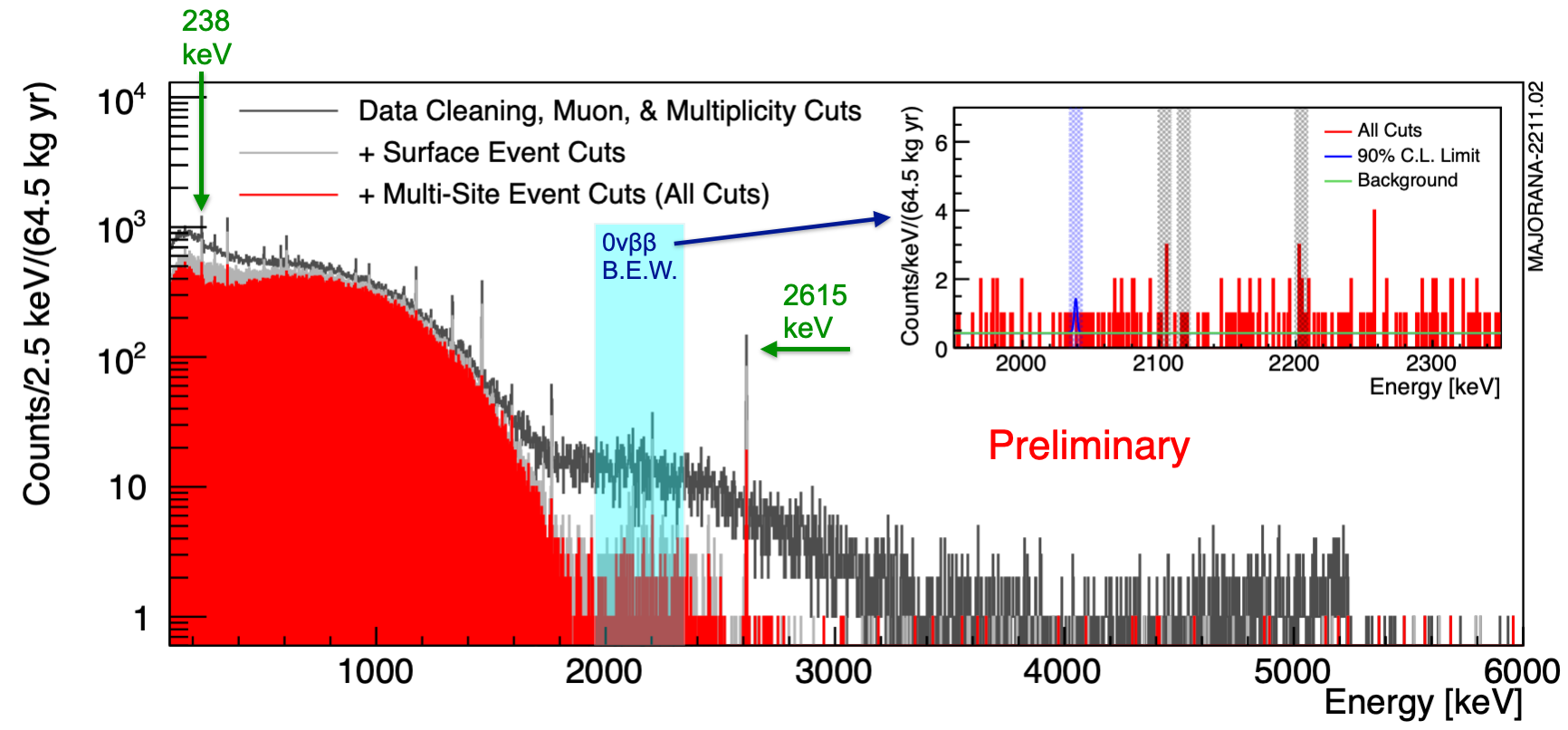}
    \caption{The energy spectra, before and after cuts, and preliminary results from 64.5 kg-yrs enriched detector operation of the \DEMnospace{}.  The background estimation window is highlighted in light blue and shown in the inset.  The blue vertical band around 2039\,keV in the inset is the 10-keV window around $Q_{\beta\beta}$, while the gray vertical bands in the inset are the 10-keV windows around peaks from the decays of \Thallium{} and \Bismuthns{}.  These four vertical bands are excluded from the BEW.  Prominent \Thorium{} $\gamma$-ray peaks are highlighted by arrows with green text. Figure adapted from the data in Ref. \cite{znbb22}}
    \label{fig:newresults}
\end{figure}

A window surrounding $Q_{\beta\beta}$ is used as a proxy to estimate the background rate.  This 360-keV-wide background-estimation window (BEW) ranges from 1950 keV to 2350 keV and 
%excludes four 10-keV-wide regions around $Q_{\beta\beta}$ and
excludes four 10-keV-wide regions: one around $Q_{\beta\beta}$, and three around 
prominent $\gamma$-peaks from the decays of \Thallium{} and \Bismuthns{} (see Figure \ref{fig:newresults}).  The assay-based background model predicts a rate of 2.9 $\pm$ 0.14 cts/(FWHM t y) in the BEW.

\subsection{\label{sec:introbmoq}Open Questions on Background Modeling}

\begin{figure}
    \centering
    \includegraphics[scale=0.35]{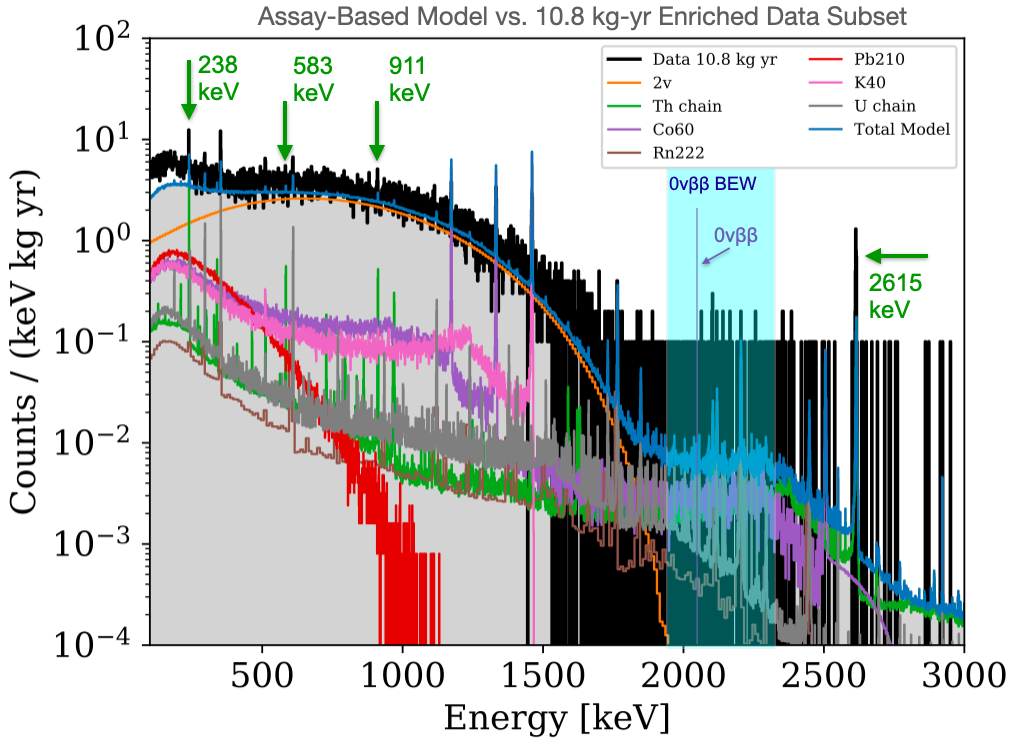}
    \caption{Predicted energy spectrum of the assay-based background model (top, blue curve) plotted against a 10.8-kg-yr  subset of the enriched-detector data (black).}
    \label{fig:assayvdata}
\end{figure}

The \MJ collaboration ended operation of the enriched \EnrGe{}detectors in March 2021.  Subsequent analysis of data from the enriched detectors produced a final background rate in the BEW of 15.7 $\pm$ 1.4 cts/(FWHM t y) from the low-background configuration spanning most of 64.5 $\pm$ 0.9 kg-yrs active exposure (see Figure \ref{fig:newresults}) \cite{znbb22}.  This illustrates the first open question that the \MJ collaboration aims to address.  Why is there more than a factor of five difference between the final background rate and the predicted rate?  The answer lies in the \Thorium{} decay chain, as shown in Figure \ref{fig:assayvdata}.  The measured rate in the prominent \Thorium{} chain $\gamma$-ray peaks, particularly at 238 keV and 2615\,keV, exceeds the predicted values by about a factor of two and five respectively.  The second open question concerns the difference in the final background rate between the two modules.  In Module 1, the rate in the same low-background configuration is 18.6 $\pm$ 1.8 cts/(FWHM t y), whereas in Module 2 the rate is  only $8.4^{+1.9}_{-1.7}$ cts/(FWHM t y).  This difference indicates that the \Thorium{} excess observed in the total rate is non-uniform and larger in Module 1.

The \MJ collaboration is addressing both of these questions through an analysis of the observed intensities of spectral lines from \Thorium{} in natural detectors based on position in Module 1, as well as by performing a fit to the energy spectra of all detectors with a background model that accommodates for the source of the excess.  The following two sections detail this progress.

\section{\label{sec:direct} DIRECT EVIDENCE OF NON-UNIFORM BACKGROUND EXCESS}
% The \DEM operates with three types of detectors, enriched p-type point contact crystals, natural BEGe crystals, and enriched inverted coaxial point contact cyrstals.  These detectors were deployed in various configurations throughout the lifetime of the \DEM and are shown in figure.  Note that at various stages, there were multiple enriched and natural detectors offline for operational reasons.  Additionally, the \DEM operated with a module of only natural detectors in 2021 after the removal of enriched detectors for the LEGEND experiment.

%\subsection{\label{sec:directrates} Excess \Thorium{} Rates in Two Module 1 Natural Detectors}

\begin{figure}[b]
    \centering
%    \begin{subfigure}{0.49\textwidth}
    \includegraphics[scale=0.24]{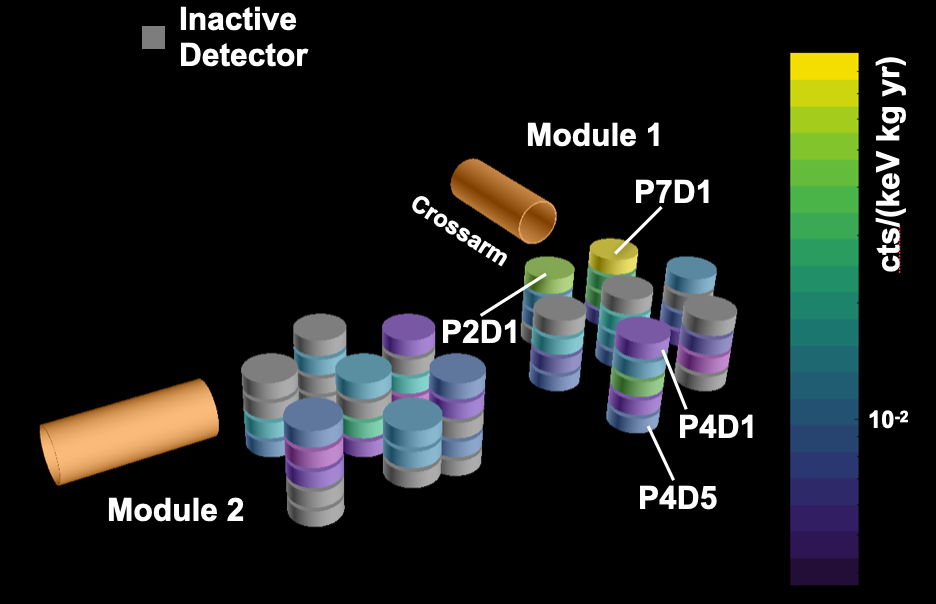}
%    \caption{Rate of the 2615 keV peak in individual detectors through datasets before the removal of enriched detectors.  Detectors in gray were not operational during this period.}
%    \end{subfigure}
%    \begin{subfigure}{0.49\textwidth}
%\hfill
\hspace{0.5cm}
    \includegraphics[scale=0.45]{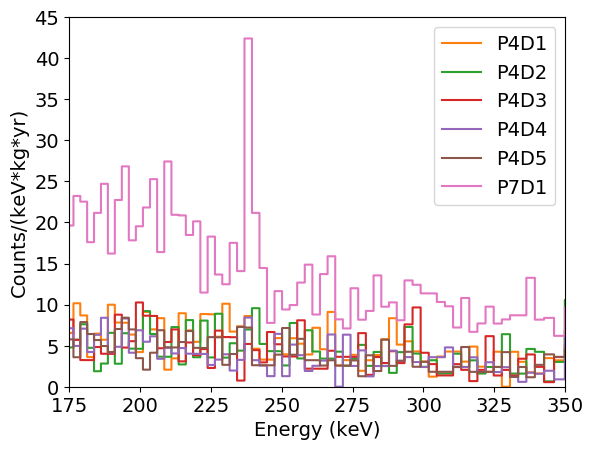}
%    \label{fig:natdet238plot}
%    \caption{Rate of the 238 keV peak in operable Module 1 natural detectors through datasets before the removal of enriched detectors.  Detector C1P2D1 is excluded due to limited statistics as it was biased down early in the dataset.}
%    \end{subfigure}
    \caption{\label{fig:natdet3dplot} \label{fig:natdet238plot} Direct evidence for a non-uniform \Thorium{} excess in Module 1 natural detectors, based on rates of \Thorium{}-chain $\gamma$-ray peaks.
     \textbf{Left:} Rate of the 2615-keV peak in individual detectors through datasets before the removal of enriched detectors.  Detectors C1P2D1 and C1P7D1 have the highest rates.  Detectors in gray were not operational during this period.
      \textbf{Right:} Rate of the 238-keV peak in operable Module-1 natural detectors through datasets before the removal of enriched detectors.  Detector C1P7D1 has a much higher rate than the other detectors.  Detector C1P2D1 is excluded due to limited statistics as it was biased down early in the dataset.
    }
\end{figure}

The detectors in each module are arranged into seven stacks, called strings (see Figure~\ref{fig:natdet3dplot} left). % 4a).  
Each string carries up to five detectors. Each detector is identified by a CPD code, where C is the cryostat number, P is the (string) position number, and D is the detector number.  The top detector (D1) in each string sits at an elevation that is just below the opening of the hollow copper crossarm where it interfaces with the cryostat.  The two strings closest to this opening are strings 2 (P2) and 7 (P7).

In datasets prior to the removal of enriched detectors for use in LEGEND, natural detectors C1P2D1 and C1P7D1 observed an excess in the 2615-keV rate compared to other natural detectors in Module 1 (see  Figure~\ref{fig:natdet238plot} left).  This suggests that a localized source of \Thorium{} exists in a component inside or near the crossarm opening of Module 1.  Furthermore, detector C1P7D1 observed an excess in the 238-keV rate compared to other natural detectors (see Figure~\ref{fig:natdet238plot} right).  The 238-keV $\gamma$-line is attenuated by more than 70$\%$ in 1 cm of copper, which suggests that minimal shielding exists in the line of sight between the source of the excess and C1P7D1.

These rates suggest that the excess either is located in the region of the crossarm opening or is integral to the crystals themselves.  When enriched detectors were removed from the \DEM in March 2021, the natural detectors from Module 1 were installed into Module 2.  C1P7D1 and C1P2D1 in particular were placed in new locations within Module 2.  When observing the integrated rate of these two detectors in datasets when they operated in Module 1 versus Module 2, their overall rate falls to a comparable level with the other natural detectors.  This measurement eliminates the possibility that the excess originates in these two detectors.

\subsection{\label{sec:directassay} New Assay Campaign}

A number of candidate sources were identified in the Module-1 crossarm-opening region.  These components are summarized in Table I.

\begin{table}
\centering
\begin{minipage}[c]{0.45\textwidth}
\includegraphics[width=\textwidth]{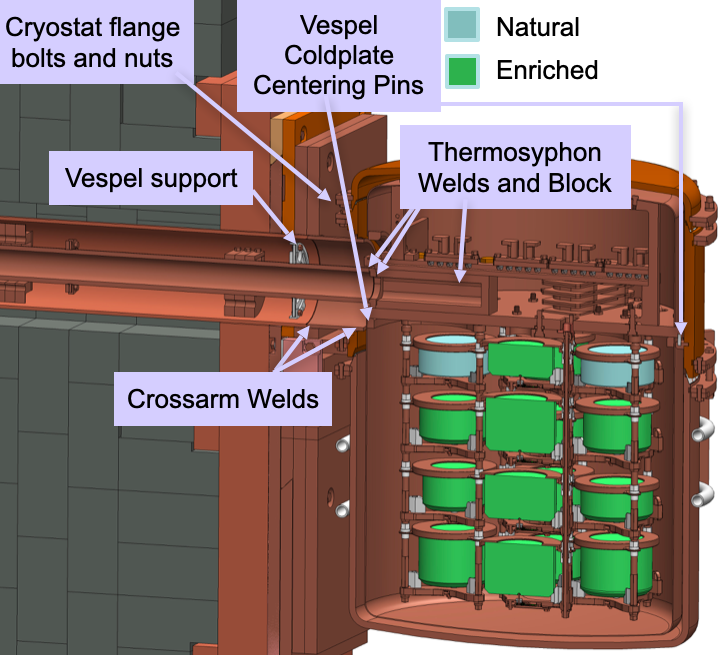}
%\caption{Graphic of the Module 1 cryostat with the locations of select candidate \Thorium{} excess components labeled.  The cables are not pictured, but both run the length of the crossarm and coil in areas in the cryostat above the detector array. }
\end{minipage}
\hfill
\begin{minipage}[c]{0.5\textwidth}
   \label{table:rearrange}
    \begin{ruledtabular}
    \begin{tabular}{lrl}
                           & Sensitivity                     & Upper Limit \\
   Module 1 Component Name & \multicolumn{1}{c}{(mBq)}       &  \multicolumn{1}{c}{(mBq)}  \\
 \hline
 Crossarm Weld 1 & $\sim$1~~~~ & ~<0.52\\
 Crossarm Weld 2 & $\sim$6~~~~ & ~<1.3\\
 Thermosyphon Welds & $\sim$4~~~~ & ~<0.72\\
 Vespel Support Ring & 3~~~~ & ~<0.09\\
 Vespel Coldplate Centering Pins & 1~~~~ & ~<0.1/pin\\
 Thermosyphon Mounting Bolts & 1~~~~ & ~~~~TBD\\
 Cryostat Flange Bolts \& Nuts & 1~~~~ & ~~~~TBD\\
 Thermosyphon Block & 1~~~~ & ~~~~TBD\\
 High Voltage Cables & 1/cable~ & <0.18/cable\\
 Signal Cables & 1/cable~ & ~~~~TBD\\
 \end{tabular}
    \end{ruledtabular}
%       \caption{Table of candidate components for the \Thorium{} excess, all from the crossarm opening region, as labeled in the Figure to the left.  Their assay sensitivity requirements and assay results are shown in the middle and right columns, respectively.}
\caption{Table of candidate components for the \Thorium{} excess, with assay sensitivity requirements and assay results.  All candidates are from the crossarm opening region, as labeled to the left.  The cables (not pictured) both run the length of the crossarm and coil in areas in the cryostat above the detector array.}
\end{minipage}%

%\begin{subfigure}{0.39\textwidth}
%\includegraphics[scale=0.25]{candidates.png}
%\end{subfigure}
%\begin{subfigure}{0.6\textwidth}
%\begin{tabular}{ |p{4.5cm}||p{1.5cm}||p{3cm}|}
% \hline
% \multicolumn{3}{|c|}{Candidate Components for \Thorium{} Excess} \\
% \hline
% Module 1 Component Name & Sensitivity (mBq) & Upper Limit (mBq) \\
% \hline
% Crossarm Weld 1 & $\sim$1 & <0.52\\
% Crossarm Weld 2 & $\sim$6 & <1.3\\
% Thermosyphon Welds & $\sim$4 & <0.72\\
% Vespel Support Ring & 3 & <0.09\\
% Vespel Coldplate Centering Pins & 1 & <0.1/pin\\
% Thermosyphon Mounting Bolts & 1 & TBD\\
% Cryostat Flange Bolts \& Nuts & 1 & TBD\\
% Thermosyphon Block & 1 & TBD\\
% High Voltage Cables & 1/cable & <0.18/cable\\
% Signal Cables & 1/cable & TBD\\
% \hline
%\end{tabular}
%\end{subfigure}
%\caption{\textbf{Left:} Graphic of the Module 1 cryostat with the locations of select candidate \Thorium{} excess components labeled.  The cables are not pictured, but both run the length of the crossarm and coil in areas in the cryostat above the detector array.  \textbf{Right:} Table of candidate components for the \Thorium{} excess, all from the crossarm opening region.  Their assay sensitivity requirements and assay results are shown in the middle and right columns, respectively.}
%\label{table:rearrange}
\end{table}

\Thorium{} decays originating in the welded-copper crossarm and thermosyphon components were simulated, and prominent $\gamma$-peaks in the simulated spectrum were scaled to data to determine the amount of activity required in these volumes to explain the excess rate.  The configuration of the \DEM after March 2021, containing only natural Ge detectors in Module 2, could be used as an assay instrument with the sensitivity to these activities, ranging between 1 -- 6 mBq.  An in-situ assay took place in which the welded parts from Module 1 were cut from their original positions and placed beside the Module-2 cryostat, within the inner copper shield, for about a month.  Analysis from this period produced upper limits on the activity for these parts that ruled them out as sources of the \Thorium{} excess.

The remaining parts in the table were removed from Module 1 and sent to $\gamma$-counting assay stations in Canada and Europe.  As of July 2022, all of these parts have either not yet completed the assay period, or have exhibited \Thorium{} rates that rule them out as sources of the excess.

\section{\label{sec:fitting} BACKGROUND-MODEL FITTING}

The fit-based background model uses simulations of \DEM components to fit the simulated energy spectra of those components to data, producing a set of activity predictions for each group of components.  Ideally, a fit of well-organized simulated component groups to data could identify the location of a \Thorium{} excess and produce the activity value(s) of such an excess.  In practice, the fit-based background model is challenged by the limited statistics of the \DEM and the complexity of the component arrangements.

\subsection{\label{sec:fittingsims} Simulations}

Monte-Carlo simulations of the \DEM are conducted in MaGe \cite{Boswell:2010mr}, a Geant4 application \cite{Agostinelli:2002hh}.  The geometry of the \DEM is composed of over 4000 parts.  Decay primaries are generated on and within each part, with a distribution that is weighted by surface area and mass respectively.  The decay primaries simulated include the primordial decay-chain elements assuming secular equilibrium (\Thoriumns{}, \Uraniumns{}, \Potassiumns{}), cosmogenic contaminants (\CosmoGe{}, \CosmoCo{}, \CosmoCotwo{}), surface contaminants (\Lead{}, \Uranonns{}), and two neutrino double beta decay ($2\nu\beta\beta$).

Studies of relative peak intensities and rates of coincident gammas in the \Thorium{} chain were carried out on energy spectra from data as well as from simulations.  When testing the hypothesis of components very close to the detectors as sources, including from the copper detector-unit parts and the low-mass front-end board that is used to readout detector signals \cite{Majorana:2021mtz}, the relative peak intensities in simulations did not match data.  Similarly, when testing the components on the far side of the hollow crossarm from the detectors as sources, including stainless steel high-vacuum-system parts, the relative peak intensities also did not match data.  These simulation studies conclude that the source of the \Thorium{} excess must exist in a middle region that includes the crossarm itself and components in the cryostat not immediately adjacent to the detectors.

\subsection{\label{sec:fittingdetails} Background-Model Fitting Details}

The \MJ collaboration has taken a two-pronged approach to background-model fitting, utilizing Frequentist and Bayesian techniques.  Both methodologies fit simulated energy spectra to data from 100 keV to above the 2615-keV peak.  Since the fit-based background model is primarily accounting for $\gamma$-ray backgrounds instead of $\alpha$-decay surface events, a cut \cite{Majorana:2020xvk} is applied to data to remove surface-alpha events.

Both fitting methodologies float 100 source activities.  These activities represent more than 4000 individual parts that are sorted into 27 component groups, based on location and material.  In turn, these 27 component groups are coupled with up to 9 decay chains each, corresponding to the isotopes listed in previous section.  Twelve independent spectra are simultaneously fit in order to produce the final set of source activities.  These spectra are categorized by simulated geometries with one or both modules present, events in which only one or more than one detector is triggered in the event window, events in which either an enriched or natural detector is triggered, and events that occur in either Module 1 or Module 2.

The Frequentist fitting algorithm utilizes a Barlow-Beeston likelihood \cite{BARLOW1993219}, itself an extension of a Poisson likelihood, in order to handle limited statistics in the simulations.  A Migrad minimizer from the Minuit \cite{JAMES1975343} python package is used, and no assay information is incorporated into the fit.  The Bayesian fitting algorithm uses a Poisson likelihood and a posterior sampler based off Hamiltonian Monte-Carlo called the No U-Turn Sampler (NUTS).  NUTS is the default sampler from the PyMC3 python package \cite{pymc3}.  Assay values are incorporated into the fit as priors, taking the form of truncated Gaussian distributions with means at the assay value and standard deviations based on the assay value error.

\subsection{\label{sec:fittingvalidation} Method Validation Using Simulated Data}
Due to the low backgrounds in the \DEMnospace{}, there are limited statistics available to inform the fit-based background model.  Therefore, both fitting algorithms must be validated against simulated datasets to evaluate their performance with low statistics.

To generate a simulated dataset, a model is constructed from 100 simulated spectra that correspond to all of the component group and decay chain tuples.  Each spectrum is weighted by a known activity density, which is typically a weighted sum of assay values from all of the components in the component group.  Data are then randomly sampled from this model to form a dataset that corresponds to a desired exposure.

When both fitting algorithms were applied to high-statistics datasets ($\sim$1000 times larger than the full exposure of the \DEMnospace{}), they performed well.  All 27 component groups fit %out 
the correct activity in every decay chain.  However, when both were used on low-statistics datasets at the level of $\sim$65 kg-yrs of exposure, they were unsuccessful in disentangling \Thorium{} contributions from various component groups, especially those with similar detector efficiencies.  Additionally the Bayesian fitting algorithm was found to be severely prior-dependent at low statistics.  These results indicated that there are insufficient statistics from the \DEM to allow these algorithms to identify a particular source or set of sources of the \Thorium{} excess, at least at the granularity level of the 27 component groups.

However, by taking the same fitting results from the Frequentist algorithm and combining each component grouping into three broad regions based on detector proximity (``near," ``middle," and ``far"), the correct \Thorium{} activity was fit.  Combining groups this way eliminated the degenerate efficiencies that hindered \Thorium{} chain $\gamma$-ray counts from being placed in the correct component.  This method of combining results allows the Frequentist fitting algorithm to place the source(s) of the \Thorium{} excess at a regional scope, corresponding to either near, far, or somewhere in the middle relative to the detector array.

\subsection{Frequentist Fit to an Open Data Subset}

As systematic studies of both fitting algorithms are ongoing, only fits to open (un-blinded) subsets of data have been performed.  One such fit was applied to a subset of open data accounting for 18.3 kg-yrs of enriched-detector exposure.  The results of this fit are shown in Figure 5.  The fit is consistent with prior results from the simulation studies of \Thorium-chain peak intensities that imply the \Thorium{} source is not from a component near the detector array or very far from the detector array.  The fit is also consistent with direct evidence from the elevated rates of two natural detectors for a non-uniform \Thorium{} excess in Module 1, as the Module-1 groups fit out a higher activity than the Module-2 groups.

\begin{figure}
    \centering
%    \begin{subfigure}{0.74\textwidth}
%    \includegraphics[scale=0.3]{freqfit.png}
   \includegraphics[width=\textwidth]{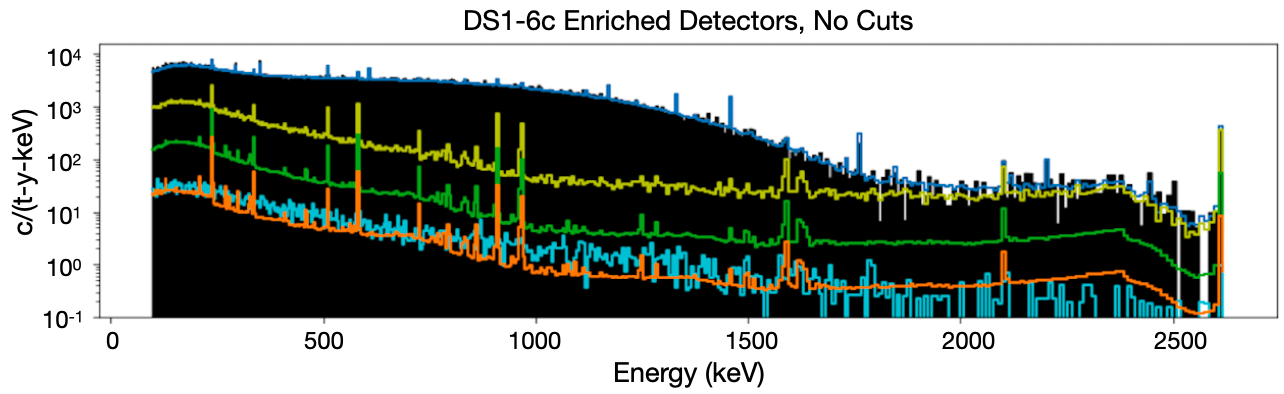}
%    \end{subfigure}
%    \begin{subfigure}{0.25\textwidth}
%    \includegraphics[scale=0.4]{freqfitlegend.png}
%    \end{subfigure}
    \label{fig:freqfit}
    \caption{Frequentist fit of \Thorium{} decays to an open subset of (18.3 kg-yrs) enriched detector data (dark filled).  The summed fitted spectrum (dark blue top curve) consists of contributions (with curves from top to bottom) from the Module-1 middle (yellow), Module-2 middle (green), far (light blue), and Module-2 near (red) component groups.  The Module-1 near group is not shown because most of its spectrum lies below $10^{-1}$ c/(t-y-keV).}
\end{figure}

This result is unsurprising as the simulations that inform the fitting are themselves consistent with direct evidence for a non-uniform \Thorium{} excess in Module 1 (see Figure 6).  A simulated spectrum from a component group outside the crossarm-opening region, such as the Inner Cu Shield source, exhibits uniform efficiencies across all natural detectors, whereas simulated spectra from a component group inside the crossarm-opening region exhibits elevated efficiencies in the two natural detectors with elevated rates in data, C1P2D1 and C1P7D1.

\begin{figure}
    \centering
%    \begin{subfigure}{0.49\textwidth}
    \includegraphics[scale=0.55]{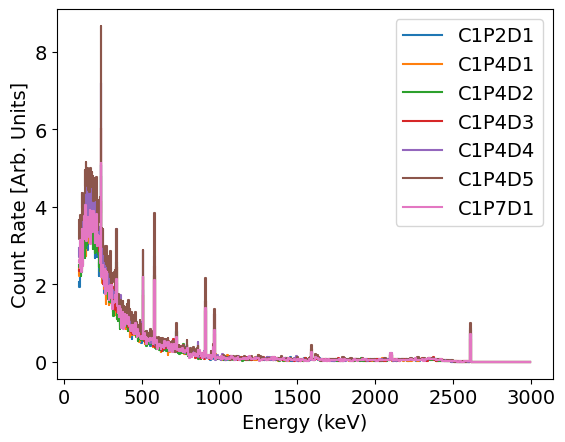}
%    \end{subfigure}
%    \begin{subfigure}{0.49\textwidth}
    \includegraphics[scale=0.55]{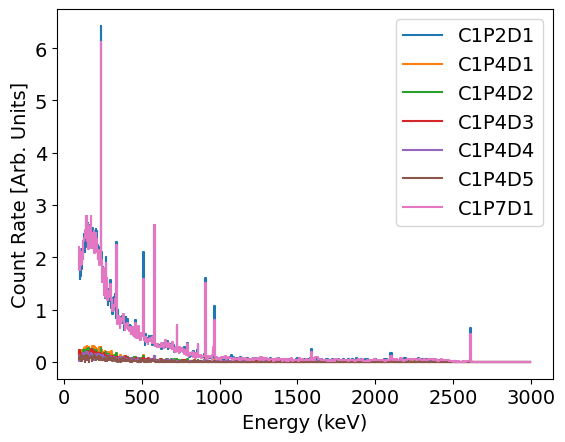}
%    \end{subfigure}
    \label{fig:sims}
    \caption{Simulated energy spectra from two of the 27 component groups in Module 1. Unlike the Inner Cu Shield source outside of the crossarm opening region, which falls into the ``far" region (left), the component group representing the crossarm opening region in the ``middle" region (right) shows higher efficiencies for C1P2D1 and C1P7D1 than other natural detectors.}
\end{figure}

\section{Conclusions}

The \MJ collaboration is addressing two open questions concerning the observed background.  The overall rate in the BEW is five times larger than assay-based predictions, and the rates differ between Module 1 and Module 2 by a factor greater than 2.  Two natural detectors adjacent to the Module-1 crossarm opening observed high integrated count rates and prominent \Thorium{} peaks compared to other natural detectors, indicating the presence of a spatially non-uniform \Thorium{} excess.  Simulations of components in this crossarm region show good qualitative agreement with data.

Method validation of the Frequentist fitting algorithm demonstrates its ability to fit at low statistics.  In the \Thorium{} decay chain, this necessitates recombining results into source regions to disentangle component contributions.  The Frequentist fitting algorithm can fit the activity of $2\nu\beta\beta$ at reasonable precision even at low statistics.  Therefore, the goal of making a measurement of the $2\nu\beta\beta$ half-life can be achieved using the fit-based background model.  These efforts are in support of constructing a complete background model of the \MJDnospace{}.  Final fits to the full set of \DEM data are underway, and new assay results for the candidate components in the crossarm region should soon be determined.

\begin{acknowledgments}
This material is supported by the U.S. Department of Energy, Office of Science, Office of Nuclear Physics, the Particle Astrophysics and Nuclear Physics Programs of the National Science Foundation, and the Sanford Underground Research Facility.
\end{acknowledgments}

\nocite{*}
\bibliography{refs}% Produces the bibliography via BibTeX.

\end{document}